\documentclass[prl,amsmath,amssymb,amsfonts,twocolumn,superscriptaddress,longbibliography,floatfix]{revtex4-1}

\usepackage{bm}
\usepackage{graphicx}
\usepackage{xcolor}
\usepackage{comment}
\usepackage{ulem}

\DeclareMathOperator{\im}{Im}
\DeclareMathOperator{\re}{Re}

\newcommand{\Nbf}{N_\text{bf}} 
\newcommand{\br}{{\bf r}}
\newcommand{\eq}[1]{\begin{align}#1\end{align}}
\newcommand{\bea}{\begin{eqnarray}}
\newcommand{\eea}{\end{eqnarray}}
\newcommand{\tcr}[1]{\textcolor{black}{#1}}
\newcommand{\DOS}{{LDoS}}

\begin{document}

\title{Multifractal correlations of the local density of states in dirty superconducting films}


\author{M. Stosiek} 
\affiliation{Physics Division, Sophia University, Chiyoda-ku, Tokyo 102-8554, Japan}

\author{F. Evers}
\affiliation{Institute of Theoretical Physics, University of Regensburg, D-93040 Regensburg, Germany}
\author{I. S. Burmistrov} 
\affiliation{L.D. Landau Institute for Theoretical Physics, acad. Semenova av. 1-a, 142432 Chernogolovka, Russia}
\affiliation{Laboratory for Condensed Matter Physics, HSE University, 101000 Moscow, Russia
}

\begin{abstract}

Mesoscopic fluctuations of 
\tcr{the} 
local density of states encode multifractal correlations in disordered electron systems. We study fluctuations of the local density of states in \tcr{a}
superconducting state of weakly disordered films. 
We perform numerical computations \tcr{in the framework of the} disordered attractive Hubbard model on two-dimensional square lattice\tcr{s}. \tcr{Our numerical results are explained by an analytical theory.} The numerical data and the theory \tcr{together} form \tcr{a} coherent picture of multifractal correlations of local density of states in weakly disordered superconducting films.  

\end{abstract}

\date{\today, v.4}

\maketitle

Superconductivity and Anderson localization are two physical effects that result in a rich variety of quantum phenomena. Initially, it was believed that 
non-magnetic disorder does not affect superconductivity (so-called ``Anderson theorem'') \cite{Gor'kovAbrikosov1959a,Gor'kovAbrikosov1959b,Anderson1959}. Later it became clear that Anderson localization in the presence of disorder can not only suppress superconductivity \cite{Sadovskii1984,Ma1985,Kapitulnik1985,Kapitulnik1986} but lead to superconductor--to--insulator transition \cite{Goldman1989} (see Refs. \cite{Goldman1998,Gantmakher2010,Sacepe2020} for a review). In the presence of Coulomb interaction even a weak disorder was predicted to be sufficients to destroy a superconducting state 
\cite{Maekawa1981,Takagi1982,Maekawa1984,Anderson1983,Castellani1984,Bulaevskii1985,Finkelstein1987,KB1993,KB1994,Finkelstein1994}. 

In the case of a short-ranged electron-electron interaction (e.g., if Coulomb interaction is screened), Anderson localization can enhance the  superconducting transition temperature, $T_c$ \cite{Feigel'man2007,Feigel'man2010,BGM2012,BGM2015}. This surprising phenomenon originates from the multifractality of electron wave functions in disordered media \cite{Evers}. Recently, theoretical predictions of Refs. \cite{Feigel'man2007,Feigel'man2010,BGM2012,BGM2015} have been \tcr{corroborated} by numerical solutions of the disordered attractive Hubbard model \cite{Andersen2018,Garcia2020}  and experimental observations of enhancement of $T_c$ in disordered niobium dichalcogenide monolayers \cite{MultifractalExp1,MultifractalExp2}. The multifractally--enhanced superconductivity is predicted to be accompanied by strong mesoscopic fluctuations of the local order parameter and the local density of states (\DOS) \cite{Feigel'man2010,Garcia2015,BGM2016,Garcia2020b,Stosiek2020,Garcia2020,Burmistrov2021}.

 Point-to-point fluctuations of the tunneling \DOS, -- frequently termed as an emergent electronic granularity, -- have been observed experimentally in many studies of disordered superconducting films \cite{Sacepe2008,Sacepe2011,Chand2012,Lemarie2013,Noat2013,Carbillet2016,Carbillet2020,MultifractalExp1,MultifractalExp2}. 
The measurements yield {\DOS} maps that are used to extract the statistics of the energy gap and the {\DOS} maximum. Recently, the spatial correlations of the energy gaps have been analysed, and the emergence of a well-defined spatial scale has been demonstrated \cite{Carbillet2016}. 

Multifractally-enhanced superconductivity was predicted to occur in a relatively narrow region in the disorder---interaction plane \cite{BGM2012,BGM2015}. 
However, significant
point-to-point fluctuations of {\DOS} 
has been reported in many experiments on superconducting films that do not demonstrate 
enhancement of $T_c$ with disorder. 
Therefore, there is a question about the origin of an emergent electronic granularity, especially in a weakly disordered superconducting films.

In this Letter we investigate numerically and with analytical means, the fluctuations of the  {\DOS} in the superconducting state of weakly disordered films. 
We perform numerical computations of disordered attractive Hubbard model on a two-dimensional square lattice. We focus on the energy dependence of the variance of the {\DOS}. 
Even at very low level of disorder we 
observe 
pronounced fluctuations of the {\DOS}.   
Our numerical findings are complemented with the analytical theory for the fluctuations of the {\DOS}. The most striking observation both in the numerics and in the theory is the logarithmic divergence of the {\DOS}-variance (at energies higher than the energy gap) with a system size $L$. 
Such divergence is a direct 
signature
of multifractal behavior of the \DOS,
originating from mesoscopic fluctuations. 
We demonstrate that 
the numerical findings and the analytical results  form a coherent picture of multifractal correlations of the {\DOS} in weakly disordered superconducting films.

\begin{figure*}[t]
	\centerline{\includegraphics[width=0.3\textwidth]{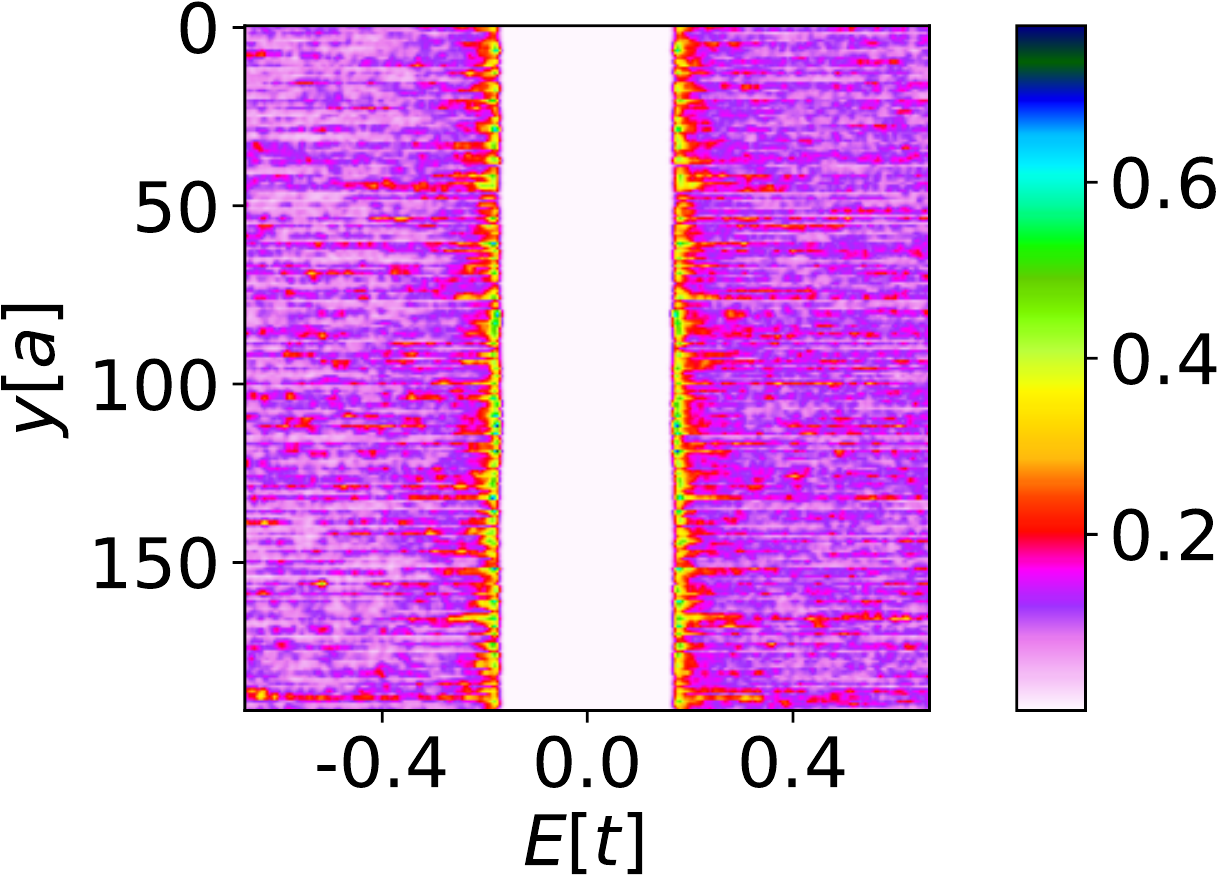}
	\quad \includegraphics[width=0.215\textwidth]{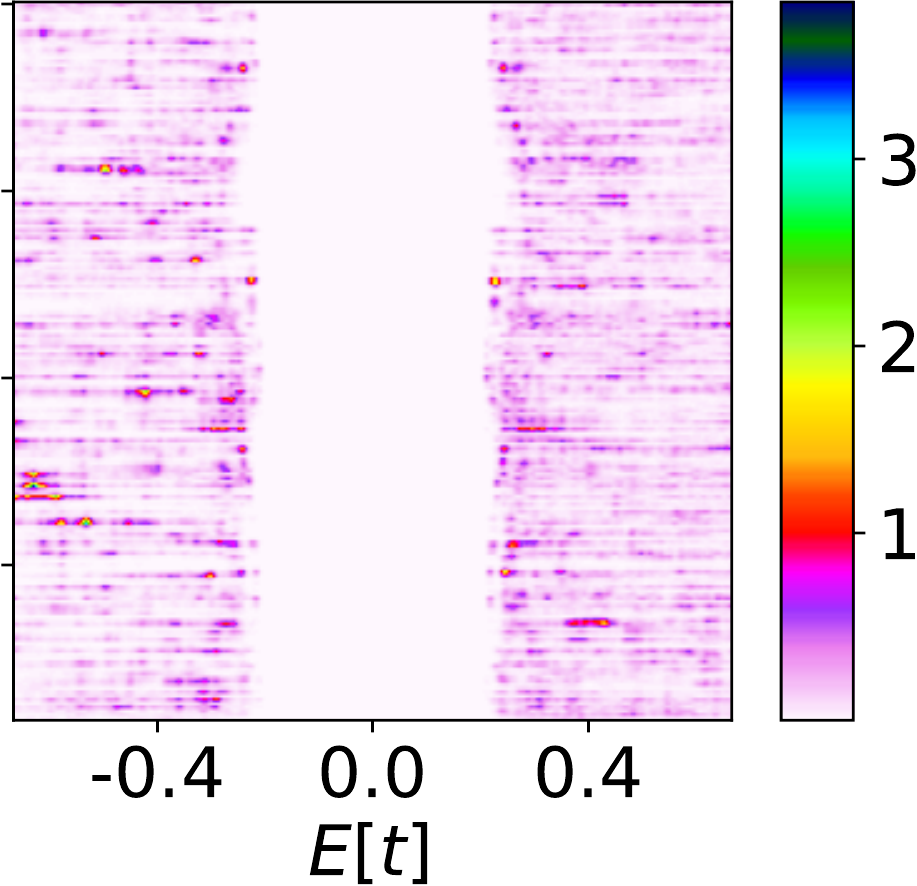}
	\quad \includegraphics[width=0.22\textwidth]{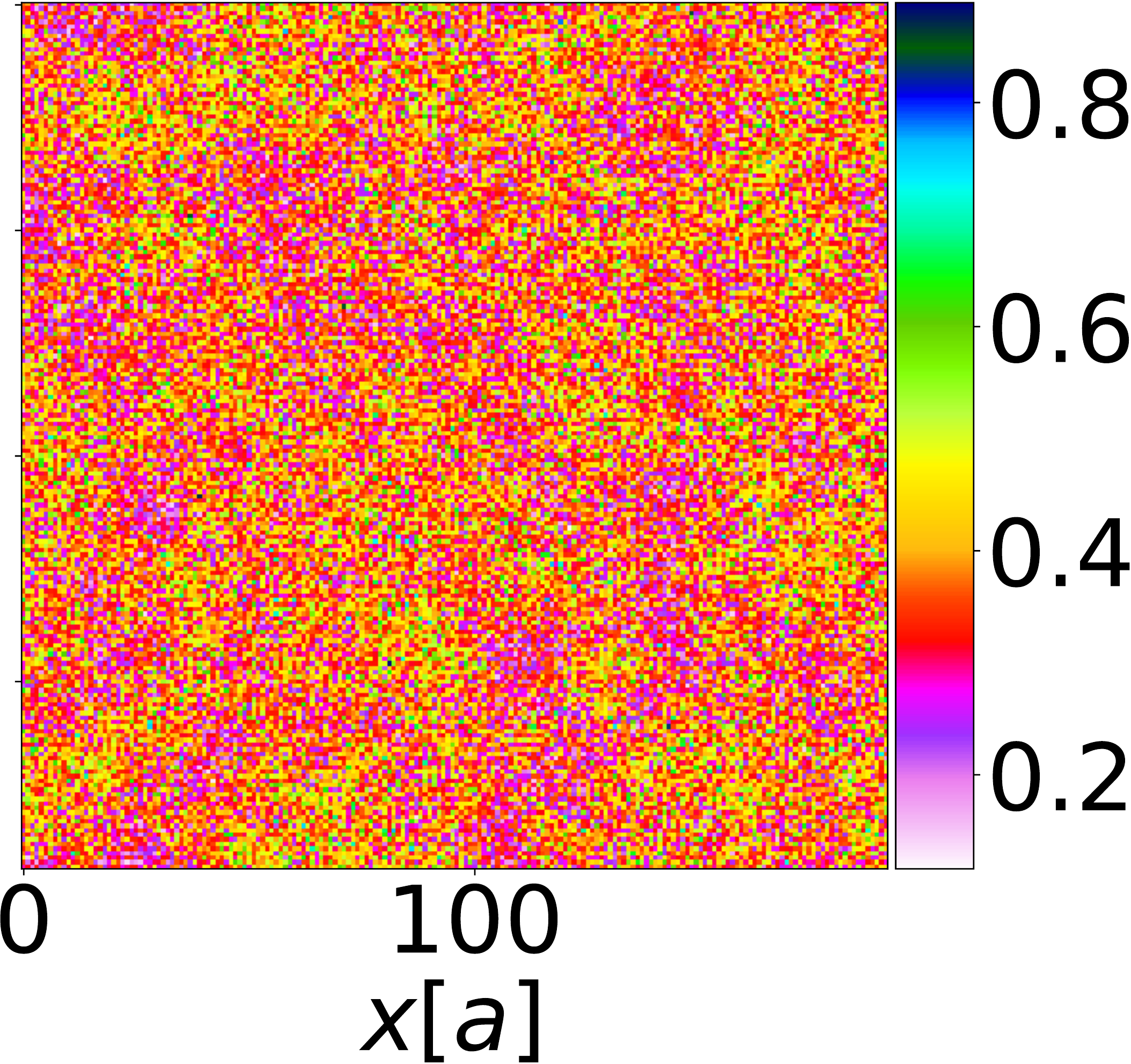}
	\quad \includegraphics[width=0.22\textwidth]{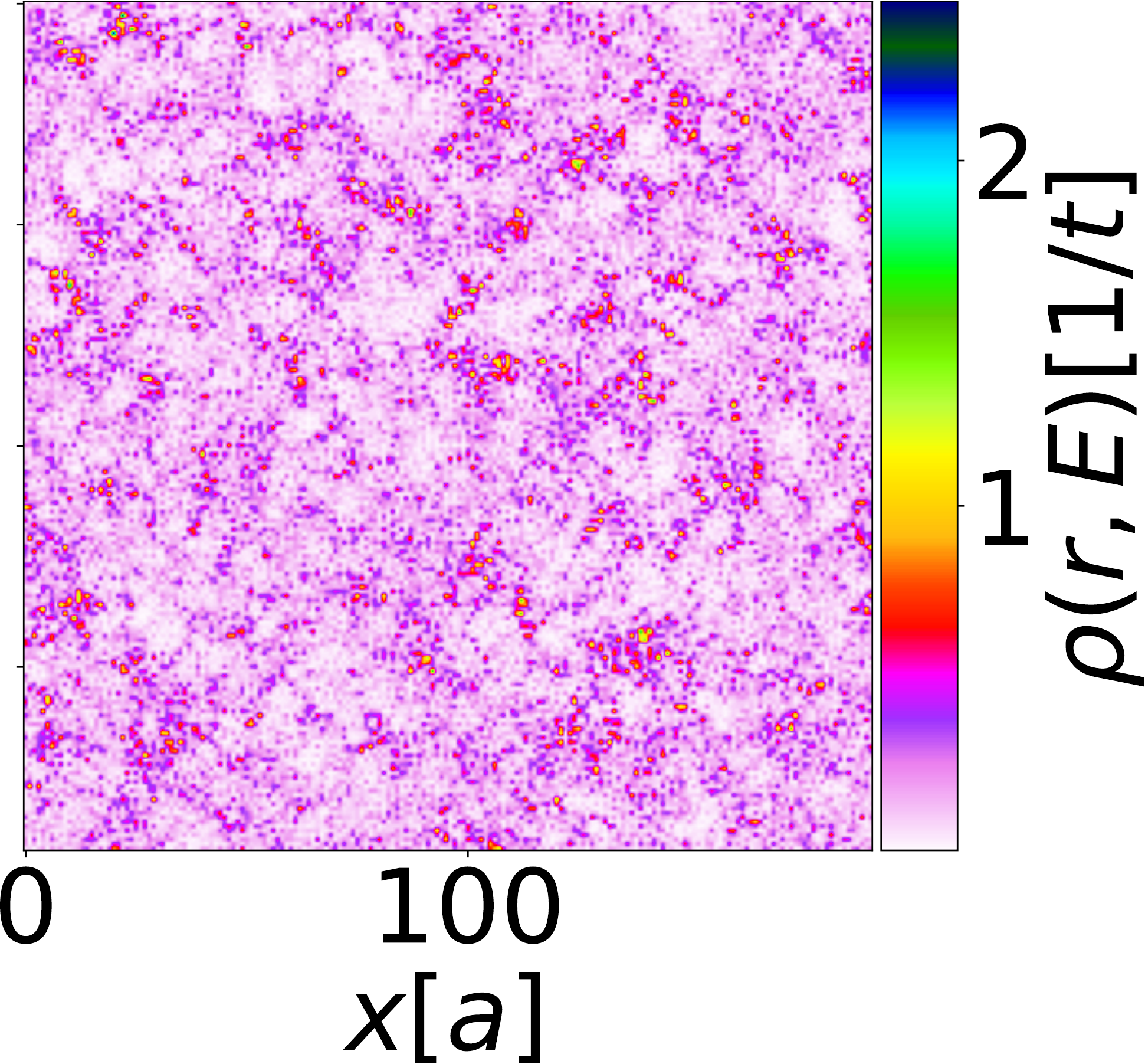}} 
	\caption{{Left Panels:} {\DOS}-linecuts { of typical samples along axial $y$-direction of the lattice and $x=0$} for W=0.5 {(outer left panel)} and W=1.25 {(middle left)}. 
	{Right panels: {\DOS}-maps of the same samples as in the left panels for $W=0.5$ (middle right) and $W=1.25$ (outer right) on the entire lattice for $E=E_\text{max}$ with the peak energy of the disorder-averaged DoS $E_\text{max}$.} 
	(Parameters: \tcr{zero temperature, 0.3 filling,} $U=2.2t$, $L=192$, and $N_\mathcal{C}=8192$.)}
	\label{Fig.Num.LocalDOSMap}
\end{figure*}

\noindent\textsc{Numerics.} We consider the attractive$-U$ Hubbard model \cite{hubbard63} on the square lattice in two-dimensions with double-periodic boundary conditions. Within the mean-field approximation the Hamiltonian reads ($U>0$)
\begin{gather}
\label{e6}
\hat H = 
-t\sum_{\langle i,j\rangle, \sigma} \hat{c}_{i,\sigma}^\dagger \hat{c}_{j,\sigma} + 
\sum_{i, \sigma} \bigl ( V_i-\mu-U n(\bm{r}_i)/2  \bigr) \hat{n}_{i,\sigma} 
\notag
\\ 
+ \sum_{i} \Delta(\bm{r}_i) \hat{c}_{i,\uparrow} \hat{c}_{i,\downarrow} + \text{h.c.}  
\label{eq:Ham:def}
\end{gather}
Here $\hat{c}_{j,\sigma}^\dagger$ and  $\hat{c}_{j,\sigma}$ stand for the creation and annihilation operators of a fermion with spin projection $\sigma=\pm 1/2$ at a site $j$. An on-site random potential is drawn from the box distribution, $V_i \in \left[-W,W \right]$. 
The chemical potential $\mu$  fix\tcr{es} the filling factor to $0.3$; throughout this work the interaction is taken as $U=2.2 t$.  
The local occupation number $n(\mathbf{r}_i)$ and the pairing amplitude $\Delta(\mathbf{r}_i)$ are determined self-consistently,
\begin{equation}
n(\bm{r}_i)= \sum_\sigma \langle \hat{n}_{i,\sigma} \rangle, \qquad  \Delta(\bm{r}_i)=\tcr{U} \langle \hat{c}^\dagger_{i,\downarrow} \hat{c}^\dagger_{i,\uparrow} \rangle , 
\label{eq:self-consistent}
\end{equation}
where $\hat{n}_{i,\sigma}=\hat{c}_{i,\sigma}^\dagger \hat{c}_{i,\sigma}$ and the average is taken with respect to the equilibrium density matrix corresponding to the Hamiltonian \eqref{eq:Ham:def} at zero temperature. We solve Eqs.
\eqref{eq:Ham:def}--\eqref{eq:self-consistent} iteratively and terminate the self-consistency cycle when $\alpha^{(n)}$, the relative change in $\Delta(\bm{r}_i)$ in iteration $n$, is at each site $\bm{r}_i$ smaller than some $\alpha$: 
\begin{equation}
    \alpha^{(n)}=\max\limits_{\bm{r}_i}\left|\frac{\Delta^{(n)}({\bm{r}_i})-\Delta^{(n-1)}({\bm{r}_i})}{\Delta^{(n-1)} ({\bm{r}_i})}\right| < \alpha.
\end{equation}
For system size $L=192$ we chose $\alpha=10^{-3}$ and for the smaller system sizes $\alpha=10^{-4}$. We employ the Kernel Polynomial Method \cite{Weisse06} to compute $n(\bm{r}_i)$, $\Delta(\bm{r}_i)$ and the {\DOS}. The \tcr{underlying expansion of the time-evolution operator in Chebyshev polynomials to order}  $N_{\mathcal{C}}$ causes Gibbs oscillations that we deal with employing the Jackson kernel
, see Ref. \cite{Stosiek2020} for further computational details.
The ensemble averaging over observables involves, typically, several hundred samples. 

The dependence of the {\DOS}  on energy, $E$, and the spatial coordinate, $y$, along 
a cut through the sample
is shown in Fig. \ref{Fig.Num.LocalDOSMap} for two values of disorder $W$. At weak disorder, $W=0.5$, (left panel in Fig. \ref{Fig.Num.LocalDOSMap})  fluctuations of the local gap are almost absent and only after an increase upto $W=1.25$ the local gap fluctuations become more visible albeit being still small (middle panel in Fig. \ref{Fig.Num.LocalDOSMap}). 

This observation is qualitatively consistent with an analytical result for the relative fluctuations of the local order parameter, $\langle (\delta \Delta(\bm{r}))^2\rangle/\langle \Delta(\bm{r})\rangle^2 =
[4/(\pi g)]\ln (\xi_0/\ell)$ \cite{Burmistrov2021}, where $g$ represents the dimensionless conductance in the normal state, $\xi_0$ the coherence length at zero temperature, and $\ell$  the mean free path. At weak disorder, i.e. large $g$, fluctuations of the local order parameter are small. With increase of disorder, i.e. decreasing $g$, the fluctuations of the local order parameter becomes more pronounced.

\begin{figure}[b]
\centerline{\includegraphics[width=0.45\textwidth]{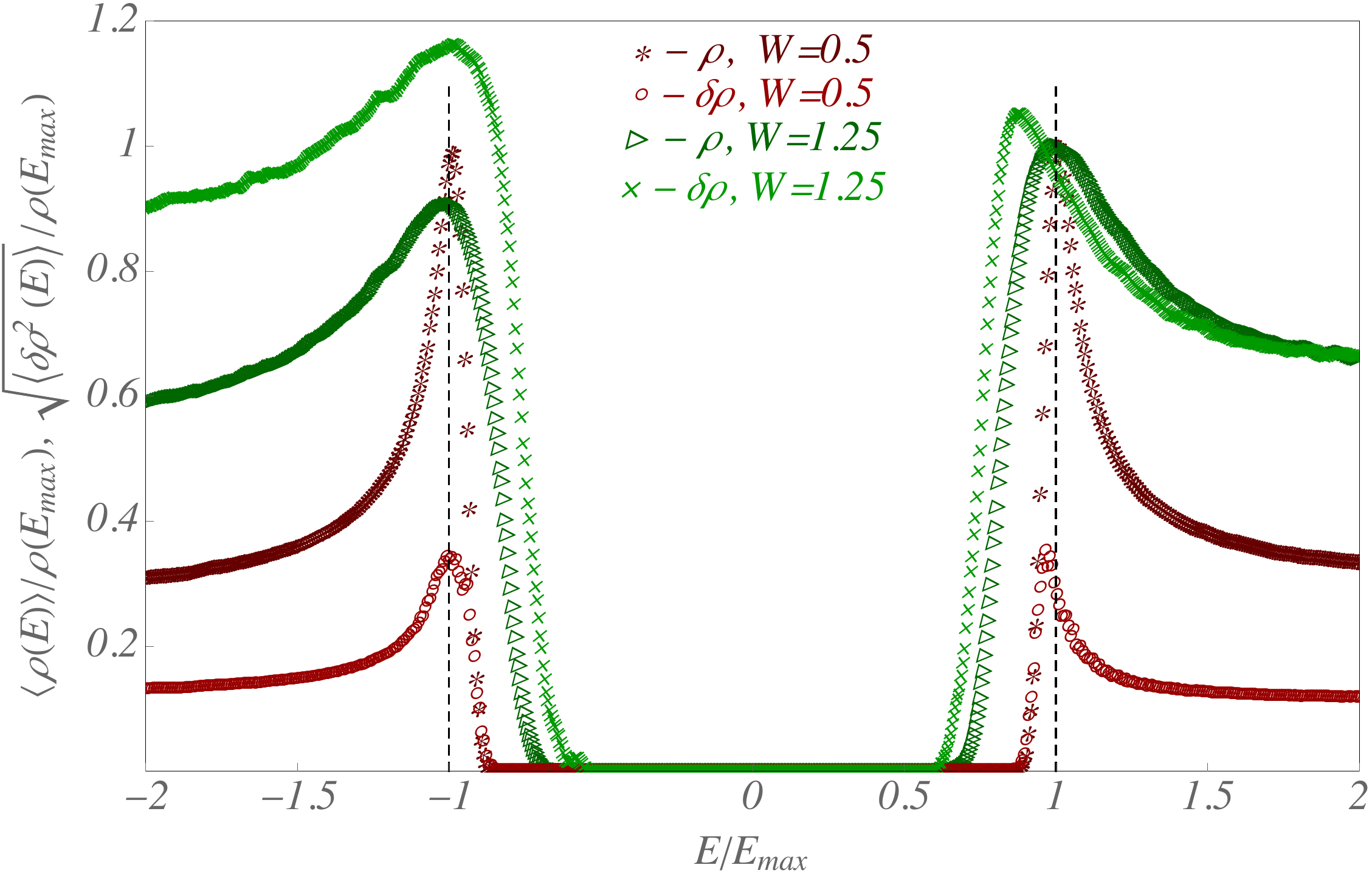}
	}
	\caption{The numerical data for the average {\DOS} and the {\DOS} variance for $W=0.5$ and $W=1.25$.
	} 
	\label{Figure:ExpDataDOS}
\end{figure}

The dependence of the average {\DOS} and its variance on $E$ are presented in Fig. \ref{Figure:ExpDataDOS} for two values of disorder. 
With increase of disorder the maximum in average {\DOS} becomes less pronounced as expected. The energy dependence of the {\DOS} variance has a form similar to the average {\DOS}, in particular, the variance has the maximum. We note that this maximum  is situated at energy $\tilde{E}_{\rm max}$ which is smaller than $E_{\rm max}$. With increase of disorder the {\DOS} variance increases. 


\noindent\textsc{Theory.} In order to understand salient features of the numerical data, we consider 2D fermions with a BSC-type attraction in the presence of a white-noise random potential, i.e. the continuum limit of the Hamiltonian \eqref{eq:Ham:def}  without the Hartree term. 
In the regime of a weak disorder, one can neglect the spatial dependence of the order parameter $\Delta$. Then the {\DOS} at a given realization of disorder can be written as 
\begin{equation}
\rho(E,\bm{r}) =\sum_{a; s=\pm} \varphi_a^2(\bm{r}) \bigl (1+\varepsilon_a/E\bigr ) \delta\bigl (E-s\sqrt{\varepsilon_a^2+\Delta^2}\bigr ) ,
\label{eq:DOS:1}
\end{equation}
where $\varepsilon_a$ and $\varphi_\alpha(\bm{r})$ are eigen energies and eigen functions of the single particle Hamiltonian in the absence of $\Delta$. Using the well-known results for statistics of eigen energies and eigen functions of weakly disordered non-interacting Hamiltonian \cite{Mirlin2000}, we compute the mean and variance of the local density of states from Eq. \eqref{eq:DOS:1} (see Supplemental Material). 

The disorder-average density of states is given as $\langle \rho(E)\rangle {=}\rho_0 \re X_E$, 
where $X_E{=}E/\sqrt{E^2-\Delta_E^2}$. Here $\Delta_E$ is the energy dependent gap function. We emphasize that such an energy dependence appears naturally in a more accurate treatment of the disordered electrons in the presence of attraction \cite{Burmistrov2021}. Although the energy dependence of $\Delta_E$ can be derived microscopically \cite{Burmistrov2021}, for a sake of simplicity, we shall use the phenomenological Dynes ansatz, $\Delta_E{=}\Delta E/(E+i\Gamma)$ with $\Gamma{\ll}\Delta$  \cite{Dynes1978}.
We note that the maximum of the average DOS is of the order of $\rho_0 \sqrt{\Delta/\Gamma}$ and is  situated at $E_{\rm max}{\simeq} \Delta+\Gamma/\sqrt{3}$. It is natural to expect that $\Gamma$ is enhanced by increasing 
disorder, so that the peak value reduces and the peak position shifts to larger values. Our numerical data, Fig. \ref{Figure:ExpDataDOS} are in qualitative agreement with this. 

\begin{figure}[t]
	\centerline{\includegraphics[width=0.95\columnwidth]{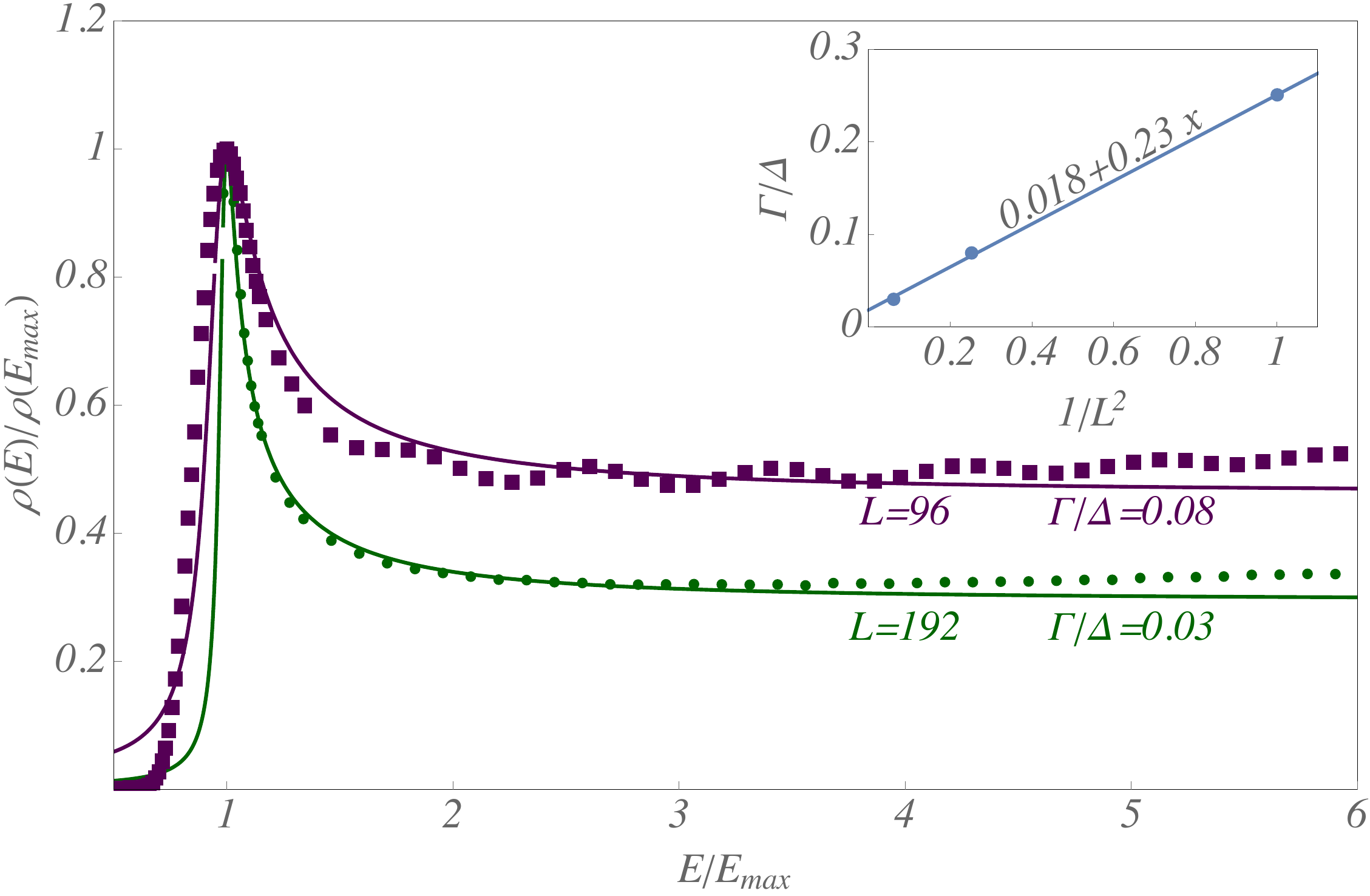}} 
	\caption{The average DOS. Comparison between numerics and Dynes expression. 
 The position of the maximum is $E_{\rm max}\approx 1$ for $L=192$ and $E_{\rm max}\approx 1.08$ for $L=96$.
 	(Parameters: $W=0.5t$, $U=2.2t$ and $N_\mathcal{C}=2048$ ($L=96$), $8192$ ($L=192$).)
	Inset: Dynes parameter $\Gamma$ versus system size for $L=48, 96, 192$. }
	\label{Fig.Num.avDOS}
\end{figure}

In order to compute the {\DOS} variance, we use the following well-known result for the irreducible part of the dynamical structure factor of non-interacting electrons in the diffusive regime (see e.g. Ref. \cite{Lee}),
\begin{gather}
\sum_{a,b} \bigl \langle \varphi_a^2(\bm{r}) \delta(E-\varepsilon_a)  \varphi_{b}^2(\bm{r})  \delta(E^\prime-\varepsilon_b)\bigr \rangle_{\rm irr} \simeq \frac{\rho_0}{2\pi}  \notag \\
\times
\int \frac{d^2{\bm q}}{(2\pi)^2}\frac{D q^2 }{(Dq^2)^2+(E-E^\prime)^2} .
\label{eq:DSF}
\end{gather}
Here $D{=}g/(4\pi \rho_0)$ denotes the diffusion coefficient.
Using Eqs. \eqref{eq:DOS:1} and \eqref{eq:DSF}, we find  the variance of the normalized local DOS at $T=0$ to the lowest order in $1/g$, 
\begin{gather}
 \sigma^2  \equiv \frac{\langle [\delta \rho(E,\bm{r})]^2\rangle}{\langle \rho(E)\rangle^2} = \frac{4}{\pi g} \re \Biggl [\ln \frac{L_E}{\ell} 
    -\frac{1+|X_E|^2}{4(\re X_E)^2}
     \ln \Bigl (\frac{E_{\rm Th}}{E}
     \notag \\+ \im \frac{1}{X_E}\Bigr )
    + \frac{1-X_E^2}{4(\re X_E)^2} \ln \Bigl (\frac{E_{\rm Th}}{E}-\frac{i}{X_E}\Bigr )
    \Biggr ] .
    \label{eq:Ldos:Var2}
\end{gather}
Here $E_{\rm Th}{=}D/(2L^2)$ stands for the Thouless energy 
and $L_E{=}\sqrt{D/(2E)}$. Also we extend the result for the variance to the energy dependent gap function ( see Ref. \cite{Burmistrov2021} for details).
Below we assume that the following condition holds $\Delta{\gg}\sqrt{\Gamma \Delta}{\gg}E_{\rm Th}{\gg}\Gamma$. This corresponds to the parameters of our numerical analysis. 

Specifically, Eq. \eqref{eq:Ldos:Var2} predicts that at  energies outside the gap-region,  $(E{-}\Delta) {\gtrsim} \Delta{\gg}\Gamma$,  the normalized variance  becomes almost  independent of $E$, and, in particular,  is logarithmically divergent with the system size,
\begin{equation}
\langle [\delta \rho(E,\bm{r})]^2\rangle/\langle \rho(E)\rangle^2 \simeq [4/(\pi g)]\ln (L/\ell) .
\label{eq:vLDOS:L}
\end{equation}
The prefactor in front of $\ln L$ in Eq. \eqref{eq:vLDOS:L}, coincides with the known result for the multifractal exponent  for a weakly disordered metal, $\Delta_2{=}{-}4/(\pi g)$ \cite{Evers}. The result \eqref{eq:vLDOS:L} implies that the spatial correlation function of {\DOS} at $(E{-}\Delta) {\gtrsim} \Delta{\gg}\Gamma$, behaves as $\langle [ \rho(E,\bm{r}) \rho(E,\bm{r^\prime})\rangle\propto  (L/|\bm{r}-\bm{r^\prime}|)^{-\Delta_2}$.
Such power-law behavior is surprising for the system with the spectral gap and the finite coherence length $\xi_0$. 

This logarithmic characteristics   is a manifestation of multifractality in the superconducting state. It can be understood on the basis of Eq. \eqref{eq:DOS:1}. To obtain the variance, each side of \eqref{eq:DOS:1} needs to be squared;
 two different contributions to the {\DOS}-variance arise: The first one correlates 
 the electron-like part of the {\DOS} ($s=+$ in Eq. \eqref{eq:DOS:1})  with the hole-like part  ($s=-$). This contribution is sensitive to the gap since the energy difference between electron- and hole-like states cannot fall below $2\Delta$. The second contribution resembles correlations between purely electron-like or hole-like states. 
The corresponding  energy difference can be arbitrarily small even in the presence of the gap and therefore fourth-order moments of wavefunction amplitudes can exhibit significant correlations that are the characteristic precursors of multifractality. 

\begin{figure}[t]
	\centerline{\includegraphics[width=0.96\columnwidth]{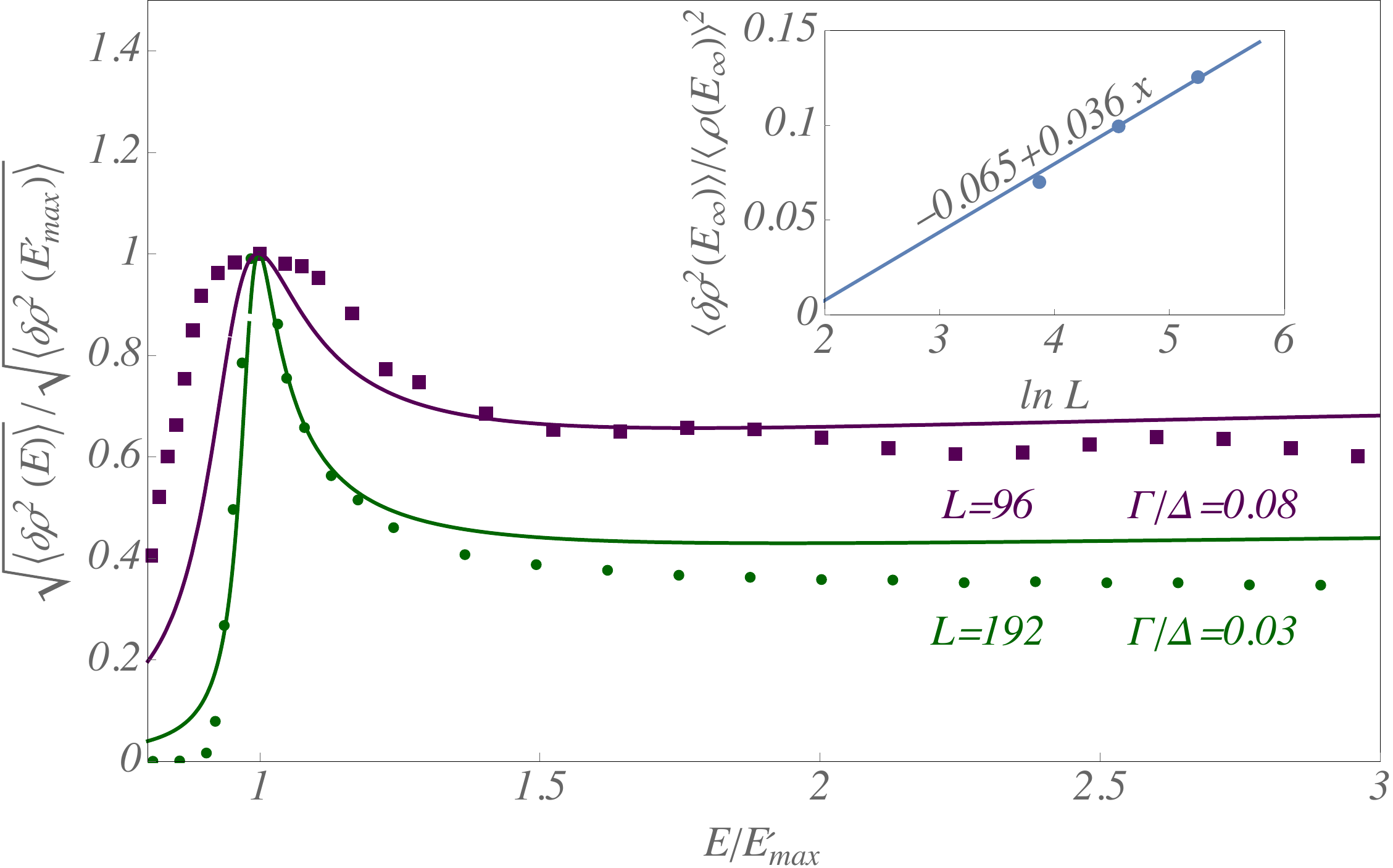}} 
	\caption{The {\DOS} variance. Comparison between numerics and theory. The parameters are the same as in Fig. \ref{Fig.Num.avDOS}. 
	\tcr{Fitting to Eq. \ref{eq:Ldos:Var2}}, 
	we \tcr{extract}   $g=35.4$ and $\ell=6$ for the fit. 
	The position of the maximum is $\tilde{E}_{\rm max}\approx 0.97$ for $L=192$ and $\tilde{E}_{\rm max}\approx 1.03$ for $L=96$.
	Inset: The {\DOS} variance at large energies, $E_\infty\simeq 3\Delta$, versus system size for $L=48, 96, 192$.}
	\label{Fig.Num.varDOS}
\end{figure}
%

Multifractality  of the {\DOS} can be seen in the average values of the higher moments of $\rho(E,\bm{r})$. In the regime of weak disorder, $g\gg 1$, the scaling of higher moments of the {\DOS} is fully determined by the second moment, $\langle \rho^q(E,\bm{r})\rangle{=}\langle \rho(E)\rangle^q (\langle\rho^2(E,\bm{r})\rangle/\langle\rho(E)\rangle^2)^{q(q-1)/2}$ \cite{Lerner1988,BGM2016,Burmistrov2021}. Therefore, using the result \eqref{eq:Ldos:Var2} we can find the distribution function for the {\DOS}. After introducing the logarithm of the normalized {\DOS}, $x=\ln[\rho(E,\bm{r})/\langle\rho(E)\rangle]$, its distribution function acquires the log-normal form, 
\begin{equation}
f(x) \approx \exp[-(x+\sigma^2/2)^2/(2\sigma^2)]/\sqrt{2\pi\sigma^2} ,
\label{eq:fx}
\end{equation}
where $\sigma^2$ is given by Eq. \eqref{eq:Ldos:Var2}.
\color{black}

\noindent\textsc{Discussions.}
In Fig. \ref{Fig.Num.avDOS}
we compare the average DOS obtained from numerical solution of Hamiltonian \eqref{eq:Ham:def} with the Dynes ansatz. We observe a reasonable agreement at weak disorder $W=0.5$  and not too small energies. We note that the Dynes parameter $\Gamma$ depends linearly on $1/L^2$ (see inset in Fig. \ref{Fig.Num.avDOS}). {This behavior is caused by our choice of the number of Chebyshev polynomials $N_\mathcal{C}$ used in the expansion of the {\DOS}}. \footnote{We choose $N_\mathcal{C}$ such that the number of states within the resulting energetic broadening is constant for all system sizes. Thus the $1/L^2$ behavior is a reflection of the dependence of the mean level spacing on system size. While there is an intrinsic broadening caused by disorder, it is negligible with respect to the broadening brought about by our numerical procedure.}


We continue the discussion with 
the {\DOS} variance. A 
detailed comparison between the numerical data \tcr{at} weak disorder $W=0.5$ and the analytical prediction, Eq. \eqref{eq:Ldos:Var2}, is presented in Fig. \ref{Fig.Num.varDOS}. 
The logarithmic growth of the {\DOS} variance with the system size agrees with numerical data as  shown in the inset to Fig. \ref{Fig.Num.varDOS}. 
Incidentally, the logarithmic dependence of the {\DOS} variance on $L$ allows us to extract values of $g$ and $\ell$. 

In agreement with numerical data, see  Fig.~\ref{Figure:ExpDataDOS}, Eq. \eqref{eq:Ldos:Var2} predicts that the DOS variance has the maximum situated at energy $\tilde{E}_{\rm max}$ which is smaller than the energy $E_{\rm max}$ of the {\DOS} maximum. In particular, one can find  $E_{\rm max} {-} \tilde{E}_{\rm max} {\propto} \Gamma/\ln (L_{\sqrt{\Gamma\Delta}}/\ell){\ll} \Gamma$. We note that the difference $E_{\rm max} {-} \tilde{E}_{\rm max}$ is enlarged with increase of $\Gamma$, i.e. of disorder. In accordance with Eq. \eqref{eq:Ldos:Var2}, the height of the maximum in $\langle [\delta \rho(E,\bm{r})]^2\rangle$ becomes of the order of $[4 \rho_0^2 \Delta/(\pi g \Gamma)]\ln (L{\sqrt{\Gamma\Delta}}/\ell)$. Again, this result is in agreement with the numerical data in 
Fig.~\ref{Figure:ExpDataDOS} in which the height of the maximum of the DOS variance is enhanced with increasing disorder. Therefore, the expression \eqref{eq:Ldos:Var2} provides reasonable description of the {\DOS} variance for a weak disorder. 

\begin{figure}[t]
    \includegraphics[width=0.96\columnwidth]{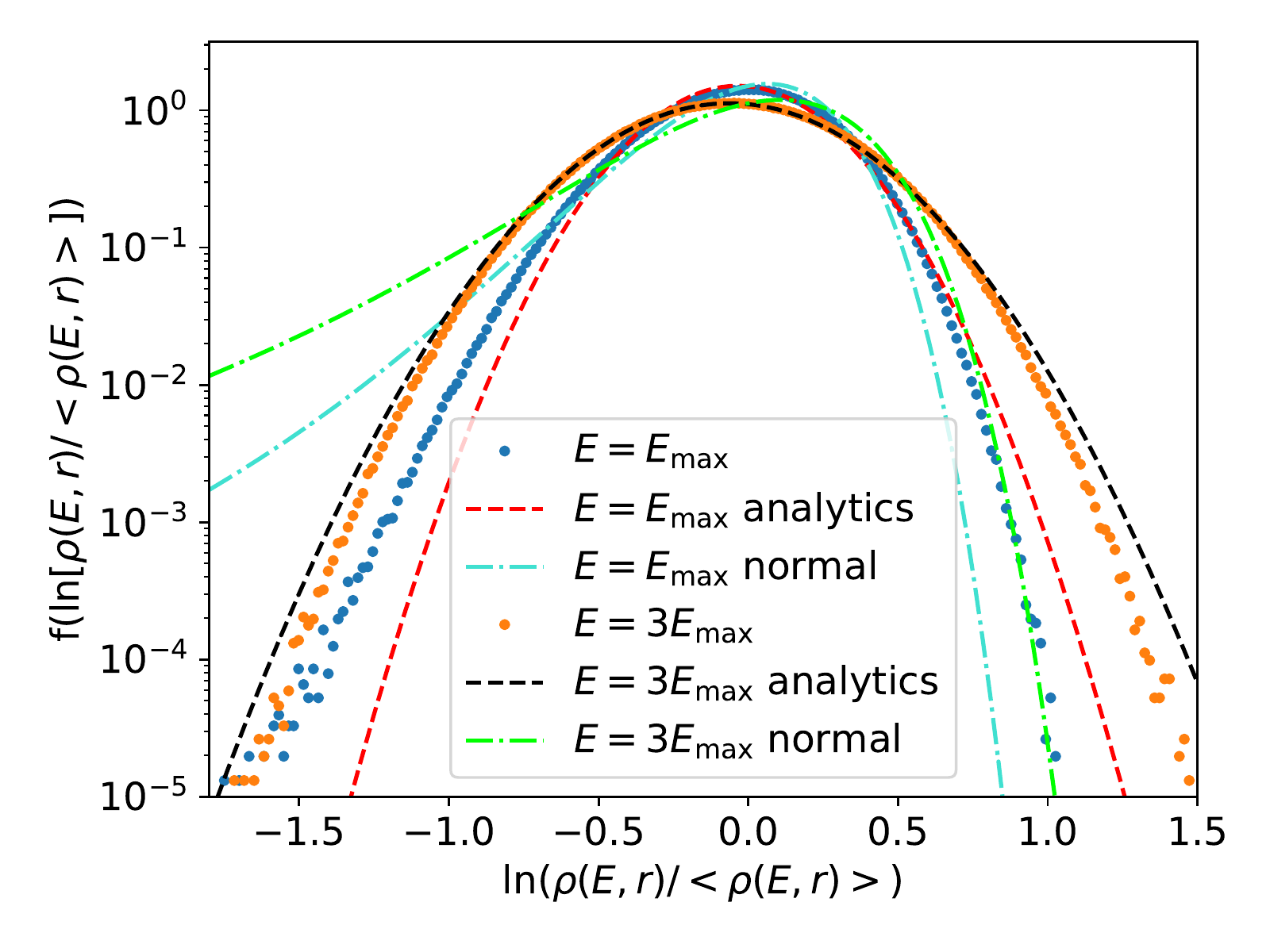}
  \caption{The distribution of the logarithm of the normalized {\DOS}, $x {=} \ln [\rho(E,\bm{r})/\langle \rho(E)\rangle]$ at two different energies $E{=}E_{\rm max}$ and $E{=}3E_{\rm max}$. The normal distribution for {\DOS} in terms of the normalized {\DOS} has the following expression, $f_{\rm n}(x){=}\exp[x {-}(e^x{-}1)^2/(2\sigma^2)]/(\sqrt{2\pi\sigma^2})$ with 
  $\sigma^2{=}\langle \delta\rho^2(E,\bm{r}) \rangle/\langle \rho(E) \rangle^2$ taken from numerics, see Fig. \ref{Fig.Num.varDOS}. (Parameters: $W{=}0.5t$, $U{=}2.2t$, $L{=}192$, 
  and $N_\mathcal{C}{=}8192$.)}
  \label{Fig5:PDF}
\end{figure}

In Fig. \ref{Fig5:PDF} we present comparison of the distribution function for the logarithm of the normalized {\DOS} taken at two energies, $E=E_{\rm max}$ and $E=3E_{\rm max}$, and obtained from numerical solution of Hamiltonian \eqref{eq:Ham:def} against the theoretical prediction of weak multifractality theory, Eq. \eqref{eq:fx}. There is reasonable agreement between numerics and the theory. Also in \ref{Fig5:PDF} we plot the curves which corresponds to normal (rather log-normal) distribution of the {\DOS} (with the same variance). As one can see, the log-normal distribution is much more in agreement with numerical data for the distribution function than the normal distribution. This supports that the fluctuations of {\DOS} seen in numerics are of multifractal origin.   
\color{black}

Finally, we mention that our numerical and analytical results are also in qualitative agreement with the experimental data \cite{private}. While the average and variance of the local DOS computed numerically as well as analytically demonstrate all features observed in the experiments, a more quantitative comparison is not indicated at this point; it would require to include, e.g., also repulsive terms into our model, which goes beyond the scope of our present work. 
We note that in experimental samples the infra-red logarithmic  divergence of the variance should be cut off by the energy-dependent dephasing length (instead of the system size) \cite{Burmistrov2021}.

\noindent\textsc{Summary.} To summarize we report the results of numerical and theoretical analysis of the energy dependence of fluctuations of the local DOS in  weakly disordered superconducting films. We found that the local DOS has 
pronounced fluctuations those variance has the energy dependence similar to the one for the average density of states. Our numerical and analytical approaches demonstrate that the DOS variance at energies higher than the energy gap diverging logarithmically 
with a system size $L$. Our  numerical findings and the analytical results make up together coherent picture of multifractal correlations of the local DOS in weakly disordered superconducting films.

We thank M. Feigel'man, K. Franke, I. Gornyi, A. Mirlin, Chr. Strunk, I. Tamir, and W. Wulfhekel for discussions. 
We are especially grateful to I. Tamir and W. Wulfhekel for providing us the experimental data on {\DOS}. The research is partially supported by the Russian Foundation for Basic Research (Grant No. 20-52-12013) Deutsche Forschungsgemeinschaft (Grant No. EV 30/11-1 and EV 30/14-1) cooperation and by the Basic Research Program of HSE.


\end{document}